\documentstyle[pre,aps,multicol,epsfig,tabularx]{revtex} 

\begin{document} 

\draft 

\title{Non-Gaussian velocity distributions in excited granular matter 
in the absence of clustering} 
\author{A. Kudrolli$^{\dagger}$ and J. Henry}
\address{Department of Physics, Clark University, Worcester, MA 01610} 

\date{\today} \maketitle 

\begin{abstract} 

The velocity distribution of spheres rolling on a slightly tilted rectangular two 
dimensional surface is obtained by high speed imaging. The 
particles are excited by periodic forcing of one of the side 
walls. Our data suggests that strongly 
non-Gaussian velocity distributions can occur in dilute granular 
materials even in the absence of significant density correlations or clustering. 
When the surface on which the particles roll is tilted further to introduce 
stronger gravitation, the collision frequency with the driving wall increases 
and the velocity component distributions approach Gaussian distributions of different widths. 
\end{abstract} 


\begin{multicols}{2}
\narrowtext

Recent simulations and experiments have shown that clusters are formed 
due to inelastic collisions in excited granular 
matter\cite{hopkins91,goldhirsh93,du95,kudrolli97,olafsen98}.
The formation of 
clusters was shown to lead to non-Gaussian velocity distributions. Here we 
report that even in the absence of cluster formation or significant 
density correlations, dissipation leads to strongly non-Gaussian velocity 
distributions.

The statistical description of dissipative systems such as granular flows are of
fundamental interest. Because thermal energies are irrelevant, the concept of
``granular temperature'' has been introduced based on the velocity fluctuations
of particles. A dissipative kinetic theory has been developed to describe rapid
granular flows\cite{campbell90}. In this theory, velocity distributions are
assumed to be locally Gaussian. Deviations from Maxwell-Boltzmann are explained 
by averaging over local Gaussian velocity distributions with different widths which
are inversely related to the local clustering\cite{puglisi99}. Given the
success of this local Gaussian distribution assumption, we might conclude
that the velocity distributions can be considered to be approximately
Gaussian for most practical situations where clustering does not occur.

To test this idea, we study the velocity distributions of steel spheres rolling inside a
two-dimensional box for which energy is supplied to the particles by
oscillating one of the side walls. A similar system has been used
previously to demonstrate the formation of clusters at high densities due to
inelastic collisions\cite{kudrolli97}. However, at low densities
corresponding to longer mean free paths and lower collision rates, the
particles appear to be randomly distributed and no clustering is observed. We
use high speed and high resolution imaging to obtain accurate velocity
distributions. Because one of the sides of
the rectangular geometry is used to input energy and the 
particle-particle and particle-wall interactions are dissipative, the
velocity distribution is
asymmetrical. An obvious consequence is that the
velocity fluctuations in the direction parallel to the driving wall 
motion is
greater than in the perpendicular direction. We find that the velocity
distribution is {\em strongly non-Gaussian even in the direction
perpendicular to the motion of the oscillating wall}. By varying the frequency
$f$ and the angle of inclination $\theta$ of the surface, we show that
non-Gaussian velocity distributions can occur when the collision rate due to
dissipative interactions is greater than the
particle-driving wall collision rate.

\begin{figure}
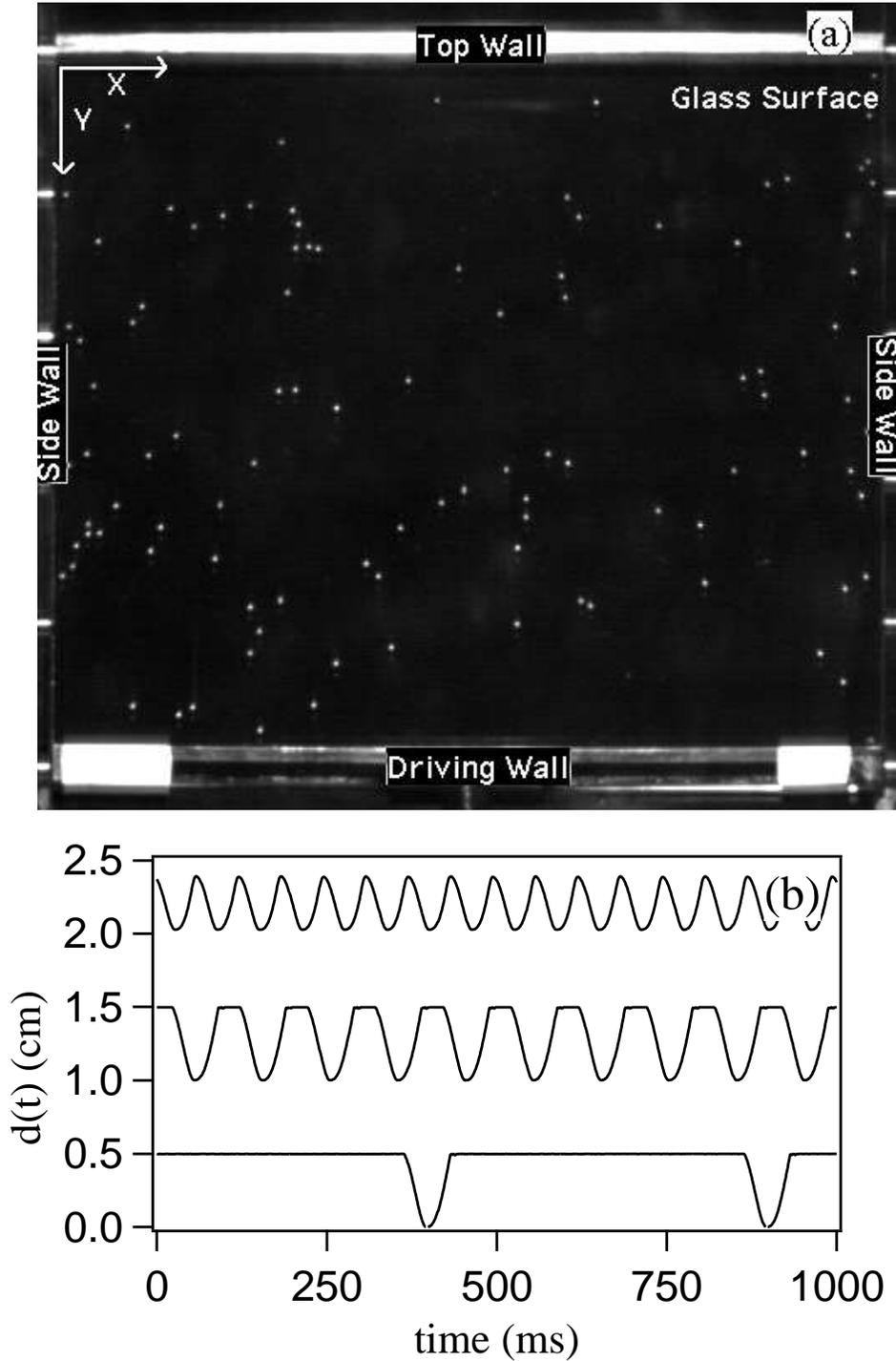

%\centerline{\epsfig{file=fig1-new.eps,width=12 cm}} 
%\centerline{\epsfig{file=piston-new.eps,width=13 cm}} 
\caption{(a) Image of 100 steel spheres rolling on a glass surface. The 
surface is tilted at angle $\theta$ to the horizontal ($\theta =
0.1^\circ$). (b) Example of driving wall
displacement $d(t)$ measured using a position sensor for frequencies $f
=2$\,Hz, 10 Hz, and 16 Hz. The displacements for the three cases are offset by
1\,cm for clarity. The wall is driven using a linear solenoid.}
\label{setup}
\end{figure}

The experimental setup consists of one hundred 0.32\,cm diameter
steel (or brass) spheres rolling on an optically smooth glass surface (29.5\,cm
$\times 24.7$\,cm) which is enclosed by steel side walls. The setup is
similar to that in Ref.~\cite{kudrolli97} with a few enhancements. The
coefficient of rolling friction of the spheres on the glass plate is less
than $2 \times 10^{-3}$. In contrast, the coefficient of sliding friction is
$\sim 0.2$. The particles are visualized using a SR-1000 Kodak
digital camera capable of taking 1000 frames per second (fps). The camera
and lights are placed such that the centers of the spheres appear bright. The
best pixel resolution of $512 \times 480$ can be obtained at 250\,fps, which
was used for most of the data presented here. A typical image is shown in
Fig.~\ref{setup}a. The particles correspond to $\sim 4$ pixels in diameter
and a centroid technique was used to find the position of the particle to
within 0.01\,cm. The particles with the highest velocities were determined
by using images separated by 4\,ms, and the velocity of the slowest particles
were determined by using images separated by at least 12\,ms. The positions
of the slow moving particles do not change appreciably in comparison to the
pixel size if a very high frame rate is used, resulting in substantial
roundoff errors. We note that a combination of both high resolution and high
frame rate is required to obtain accurate velocities. 

The driving wall is connected to a linear solenoid which is controlled by a
computer, and the resulting motion of the piston is measured with a
displacement sensor. The displacement of the piston as a function of time,
$d(t)$, for three typical driving frequencies is plotted in
Fig.~\ref{setup}b. A square waveform of fixed amplitude (4\,volt) and various
frequencies is used to obtain the displacements.

\begin{figure}
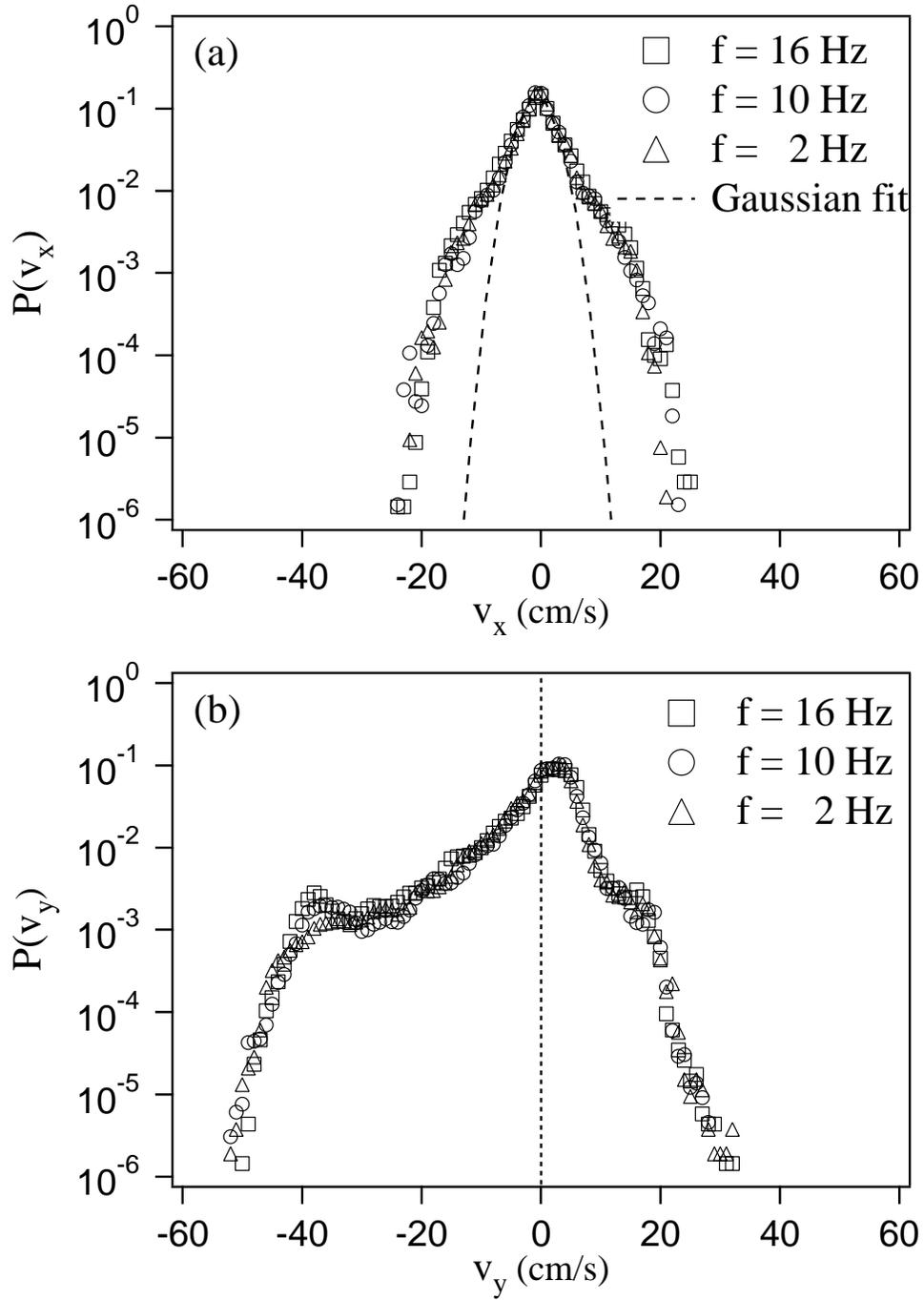

%\centerline{\epsfig{file=fig2-new.eps,width=14 cm}} 
\caption{ The probability distribution of velocity components $v_x$ 
and $v_y$ ($\theta = 0.1^\circ$). The velocity distributions are strongly
non-Gaussian. Note the peak at $v_y = -35\,{\rm cm\, s^{-2}}$ due to the
driving wall. The corresponding driving wall displacement $d(t)$ versus time
is shown in Fig.~\ref{setup}b.}
\label{vel-dist}
\end{figure}

By averaging over at least $10^5$ images, we obtain velocity distributions
of the particles rolling inside the rectangular box. The distributions of
the velocity components, $P(v_x)$ and $P(v_y)$, are plotted in
Fig.~\ref{vel-dist}a and~\ref{vel-dist}b respectively. The directions 
$x$ and $y$ and the origin are indicated in Fig.~\ref{setup}a. The surface is
slightly tilted ($\theta = 0.1^\circ$) to continously excite the particles.
Particles that are less than 2\,cm from the side walls are excluded to avoid
direct effects of the boundaries. We checked that if a narrow region in $y$ is
used, the form of the velocity distributions is independent of $y$. The plots
in Fig.~\ref{vel-dist} correspond to various driving frequencies. Note that
the form of $P(v_x)$ and $P(v_y)$ does not depend strongly on the
driving frequency $f$. The distribution $P(v_x)$ is symmetrical about 
$v_x=0$,
but is strongly non-Gaussian as can be seen from the poor fit to a Gaussian
which is also shown in Fig.~\ref{vel-dist}a. 

The distribution $P(v_y)$ shown in Fig.~\ref{vel-dist}b is non-Gaussian
and asymmetrical. The reason for the asymmetry is not difficult to 
understand. The energy is supplied to the system by the driving 
wall. Therefore, particles move with higher velocities 
in the negative-$y$ direction (indicated by the bump at $v_y \sim
-35$\,cm/s). The second more subtle reason for the asymmetry
is that the particles suffer an inelastic collision either with the
wall opposite to the driving wall or with other particles before moving in the
positive 
$y$ direction. Therefore, particles on average move with lower velocities
in the $+y$ direction than in the $-y$ direction.
Consequently, the particles also spend a longer time interval traveling in
the $+y$ direction than in the $-y$ direction. Thus,
asymmetrical $v_y$ distributions are obtained by averaging over time. Such
effects were also observed in the distributions obtained numerically by
Luding et al.\cite{luding99}. The form of
$P(v_x)$ is relatively independent of the energy input, and therefore its
non-Gaussian behavior is surprising.

\begin{figure}
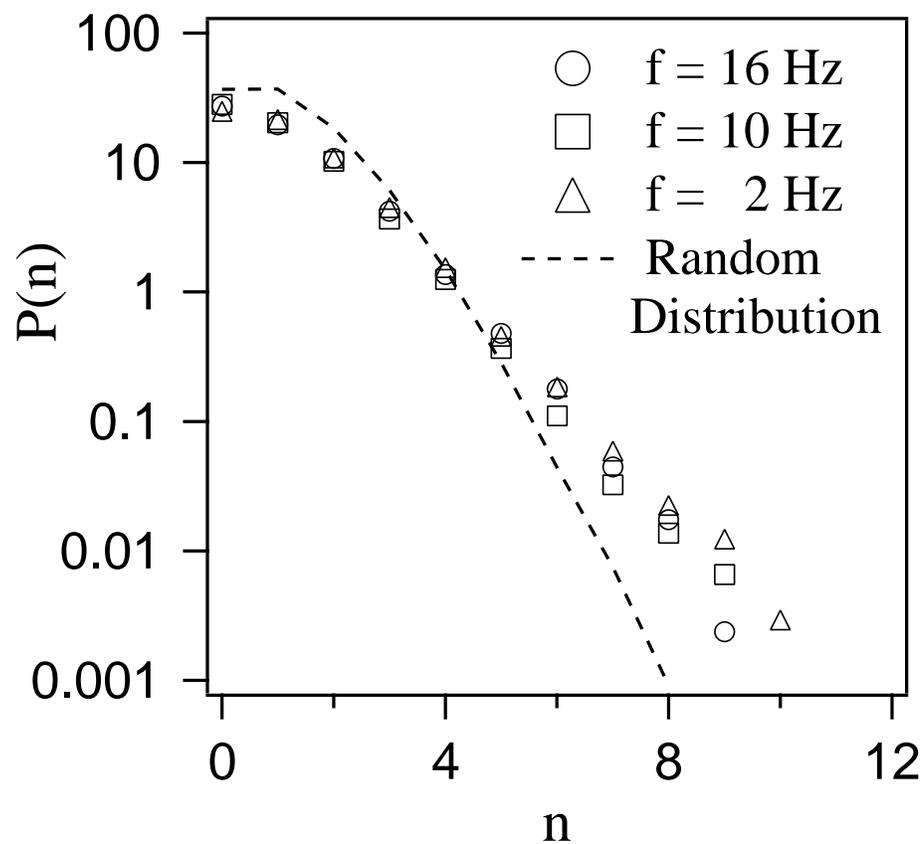

%\centerline{\epsfig{file=fig3-new.eps,width=14 cm}} 
\caption{The distribution of particles $P(n)$ in a 3\,cm $\times$ 3\,cm area.
The distribution corresponding to randomly distributed particles is also
plotted. No significant correlation in the distribution is observed.}
\label{density}
\end{figure}

In Fig.~\ref{density}, we plot $P(n)$, the distributions of the number of
particles $n$, in a 3\,cm $\times$ 3\,cm area averaged over the box and over
all images. For comparison, the distribution of randomly
distributed particles is also shown (dashed curve). Our data for $P(n)$ is
similar to the randomly distributed case with a small deviation at higher
$n$. Because of the dilute nature of our experiments, no effects of
clustering are observed as reported in Ref.~\cite{kudrolli97}.

\begin{figure}
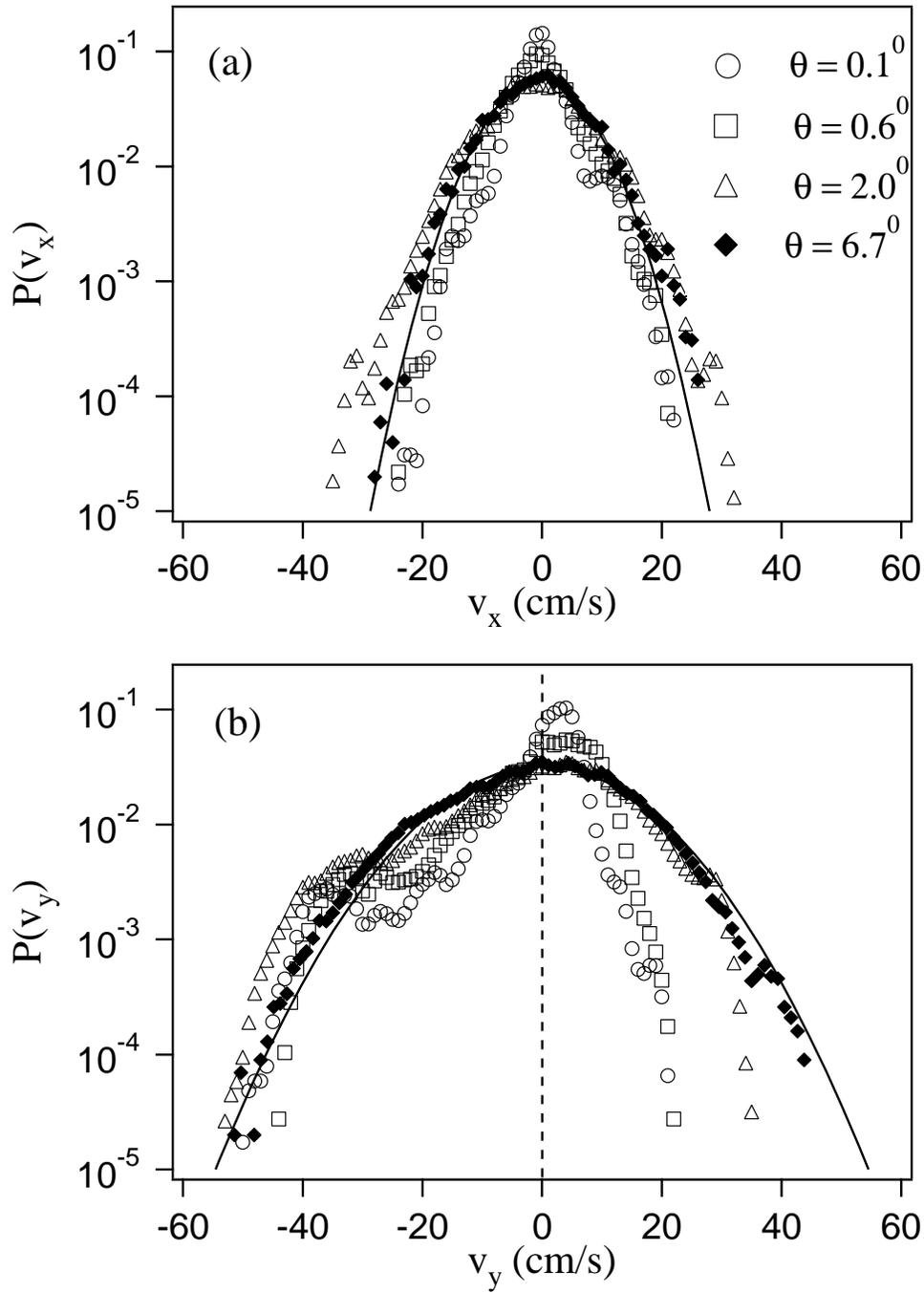

%\centerline{\epsfig{file=fig4-new.eps,width=14 cm}}
\caption{(a) and (b) Velocity component distributions as a function
of inclination angle $\theta$ ($f = 10$ Hz). All particles are in a 2\,cm
wide region at a distance $y = 16$\,cm from the top wall. The $v_x$ and
$v_y$ distributions approach Gaussians as $\theta$
is increased. The curves are a Gaussian fit for $\theta=6.7^\circ$.}
\label{pv-var}
\end{figure}

The effect of gravity and the collision rate with the driving wall is studied
by lifting the surface on which the particles roll by the tilt angle
$\theta$. The
$v_x$ and $v_y$ distributions are plotted in Fig.~\ref{pv-var} corresponding
to four values of $\theta$. The density distribution $P(n)$ is no longer
uniform in $y$ as in the $\theta=0.1^\circ$ case, but decreases strongly with
distance from the driving wall. (At the highest angle $\theta = 6.7^\circ$,
the particles do not reach the top wall.) The velocity distributions
correspond to particles in a 2\,cm wide region at a distance $y = 4$\,cm
from the driving wall to avoid the effects of the increase in potential
energy and the decrease in the average
$v_y$ at higher $\theta$. The distributions $P(v_x)$ and $P(v_y)$ grow broader 
as $\theta$ increases. A Gaussian
fit to the data corresponding to $\theta=6.7^\circ$ is shown in
Fig.~\ref{pv-var}. The data for
$P(v_x)$ is reasonably described by a Gaussian especially at low $v_x$.
However, $P(v_y)$ deviates systematically from Gaussian because of the asymmetry.
Furthermore, the width of $P(v_y)$ is large compared to the
width of $P(v_x)$ because energy is mostly supplied to the $x$
component in an inelastic collision consistent with earlier observations of
vertically vibrated granular matter\cite{warr95}. We further note that
$P(v_x)$ averaged over all $y$ is non-Gaussian for
$\theta = 6.7^\circ$ because of the slight increase in the widths of 
the Gaussians as a function of $y$. This might also be the reason for deviations from
Gaussian observed in velocity distributions of vertically vibrated monolayer layer of
particles\cite{losert99}.

As $\theta$ increases, the acceleration in the $y$ direction
increases. Therefore the
particles fall back to the driving wall faster as $\theta$ is increased. This
results in higher collision rates of particles with the driving wall leading
to higher average velocities. Thus as the collision rate with the driving
wall increases, the distributions become broader and more Gaussian.
Furthermore, the total number of inelastic collisions the particles undergo
before returning to the driving wall decreases because fewer particles hit
the top wall at higher $\theta$. 

We note that there is no discrepancy with the observation of the broadening
of the velocity distributions with an increase in $\theta$ and the
data plotted in Fig.~\ref{vel-dist}. It might appear that changing the
frequency of the driving piston should also increase the collision rate with
the bottom wall and hence have the same effect. However, the distributions
plotted in Fig.~\ref{vel-dist} do not depend strongly on the driving
frequency. The lack of the strong dependence of $P(v_x)$ and $P(v_y)$ on
$f$ in Fig.~\ref{vel-dist} can be explained by the fact that fewer particles
are struck by the forward moving wall at higher $f$. Thus the net
energy input (and collision rate with the bottom wall) appears to not depend
strongly on $f$. 

Next we discuss the effect of rolling on the distributions. In a previous
study, it was noted that rolling appears to lead to a lower effective
coefficient of restitution $r$ during collisions\cite{kudrolli97}. Some
theoretical work has been done studying rolling and
collisions\cite{kondic99}. In Fig.~5a, we plot the $r$ as a function of scattering angle $\phi$ of a rolling
steel sphere on a glass surface colliding with steel boundary walls. The
coefficient of restitution is calculated as the square root of the ratio of
the final to the initial kinetic energy. The data
is distributed over a range of $r$ which becomes smaller for higher $\phi$.
The large fluctuations are not related to measurement noise nor can they
be accounted for by variations in the initial velocity (see Fig.~5b.) 

The mean value of $r$ depends on $\phi$ and is lower for higher 
$\phi$ (indicated by the dashed line in Fig.~5a.) The mean value of $r$ 
during a wall collision averaged over all $\phi$ was $0.5 \pm 0.02$. 
In contrast the average loss of energy during a particle-particle collision
was lower and the mean $r$ was found to be 0.65 by averaging over 20
collisions. In comparison, the coefficient of restitution of the steel
sphere bouncing off a steel block after being dropped is
0.93. The larger losses in kinetic energy occur mostly
because the particles slide as their angular velocity changes
to match their new translational velocity. The change in angular
velocity was observed to occur within a centimeter even at the highest
velocities observed. It may be possible to approximately model the
complications due to rolling by simply assuming a lower
value of $r$. However, an event-driven simulation using only a
loss of kinetic energy in the normal direction does not produce non-Gaussian
distributions in $v_x$, but yields asymmetric distributions for
$v_y$\cite{tobochnik00}. The subtle effects introduced by rolling may
 help us understand the effect of higher dissipation on velocity
distributions.

Limited experiments were also performed with brass spheres. The distributions
obtained were identical to those corresponding to steel spheres under the
same conditions. Thus the velocity distributions obtained are not very
sensitive to the details of the surface properties of the spheres. 

The distribution of the speed of the particles is a
combination of the non-Gaussian or approximately Gaussian
velocity distributions of different widths (such as in
Figs.~\ref{vel-dist} and
\ref{pv-var}) and deviates from the Maxwell-Boltzmann form. Subtle
effects on the velocity distributions such as the scaling of the temperature
(the second moment of the distributions) in the near elastic limit has been
studied theoretically by Kumaran\cite{kumaran98}, Grossman et
al.\cite{grossman95}, and experimentally by Warr et al.\cite{warr95}.
However, the deviations observed in our experiments are stronger as the loss
of energy during a collision is significant compared to the energy of the
particle. Therefore scaling around the Maxwell-Boltzmann distributions
as discussed in Refs.~\cite{warr95,kumaran98,grossman95} for $r \lesssim 1 $
cannot be applied to our data.

In conclusion, our experiments on excited dissipative
granular matter rolling and bouncing inside a rectangular box have yielded
velocity distributions that are strongly non-Gaussian in the absence of
significant clustering and density correlations. Our data suggests that
caution has to be used in assuming Gaussian velocity distribution and
Maxwell-Boltzmann distributions even for rapid granular flows where position
correlations are negligible.

We thank J.\ Tobochnik and H.\ Gould for many useful discussions, and J.\
Norton for technical support. This work was supported by the donors of the
Petroleum Research Fund, and Grant No.\ DMR-9983659 from the National Science Foundation. 
A.~K.\ thanks the Alfred P.\ Sloan Foundation for
its support.

\begin{figure}
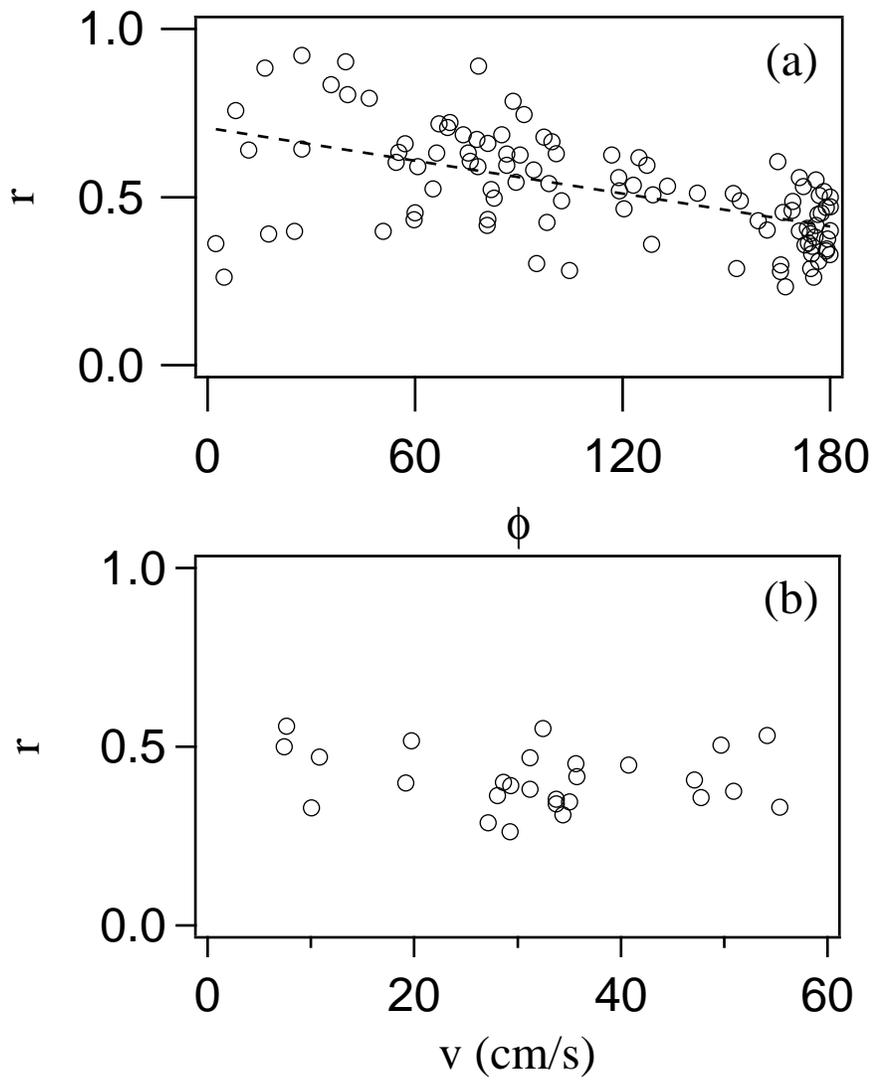

%\centerline{\epsfig{file=rest.eps,width=13 cm}}
\caption{(a) Effective coefficient of restitution $r$ versus scattering 
angle $\phi$ for a rolling particle colliding off the side walls. The dashed
line  is a least squares fit to the data. (b) The plot of $r$ versus initial
velocity
$v$ for
$170 \leq
\phi \leq 180$  fluctuates significantly.}
\label{rest}
\end{figure}

\end{multicols}

\end{document}